\documentstyle[pra,aps]{revtex}
\begin{document}
\twocolumn
\title{Entropy of phase measurement:
\\ Quantum  phase via quadrature  measurement}
\author{Zden\v ek Hradil,  Robert My\v{s}ka \cite{SLO},
Tom\'{a}\v{s}  Opatrn\'{y}\cite{KTF}, and Ji\v{r}\'{\i}  Bajer
}
\address{
 Department of Optics,
Palack\'y University \\ 17. listopadu 50,
772 07 Olomouc, Czech Republic }
\date{}
\maketitle

\begin{abstract}
The content of phase information of an arbitrary phase--sensitive
measurement is evaluated  using the maximum
likelihood estimation.   The phase distribution is characterized
by the  relative entropy--a  nonlinear functional of input quantum
state. As an explicit example the multiple measurement of
  quadrature operator is interpreted
as quantum  phase  detection achieving the ultimate resolution
predicted by the Fisher information.
\end{abstract}


 There are many approaches addressing the problem of quantum
phase measurement nowadays.  Besides the purely  theoretical
phase  concepts
anticipating the existence  of quantum phase as an  observable conjugated
canonically to the number (or difference number) operator, there
are several operational  treatments addressing the problem of
phase shift measurement within the  quantum mechanics.
Particularly,  two methods how to derive the phase information
from the phase sensitive measurement of quadrature operator
have been proposed recently.
The former one, so called ``phase (measurement) without phase
(states)"
was
formulated by Vogel and Schleich \cite{VS91}.  The method is
motivated by the  geometrical meaning  of the
quadrature-- and ideal phase  measurements  in phase space.
 The quadrature eigenstates
rotated by an angle  are used to define a phase distribution of a
single mode  of  the radiation field: A balanced
homodyne--detection  scheme measures the electric  field--strength
(variable x) probability
\begin{equation}
p(x,\theta) = |\langle \psi |x \rangle_{\theta}|^2
\label{data}
\end{equation}
in dependence on the actual  phase of local oscillator $\theta.$
This quantum detection may be interpreted as measurement
of rotated quadrature operator
\begin{equation}
 \hat X(\theta) = \frac{1}{\sqrt{2}} [ \hat a e^{- i\theta} +
\hat a^{\dagger} e^{ i\theta} ].
\label{quadr}
\end{equation}
The probability of finding  zero electric field plotted versus
local oscillator phase $\theta$
\begin{equation}
P(\theta)    =  |\langle \psi |x = 0 \rangle_{\theta}|^2
\label{vo}
\end{equation}
 constitutes the proposed phase
distribution on the interval $[0,\pi ).$
 The  phase sensitive data (\ref{data}) resulting from the
homodyne detection have been   interpreted in different  way
 by Beck, Smithey and Raymer \cite{BSR93}. Using the optical
homodyne
tomography method \cite{VR89}, the density matrix may be
reconstructed  and  represented  in the phase space.
Particularly, the authors used the representation
by Wigner function $ W(x,p)$  and linked the
phase distribution
to the
marginal   distribution  of Wigner function
\begin{equation}
P(\phi) =  \int_{0}^{\infty} r\; dr  W(x = r\cos\phi, p = r\sin\phi).
\label{faze}
\end{equation}
The resulting  phase distribution is then  periodic on the interval
$[0,2\pi).$
 Nevertheless, such an approach is suffering by
formal flaw. Since the ``probability
distribution"  (\ref{faze}) yields  negative values for superposition
of coherent states (so called ``Schr\"{o}dinger cat--like
states") \cite{GK94}, the corresponding operator measure is not  positively
defined.  The procedure  cannot be therefore
 interpreted as  any generalized  measurement \cite{Hel74}.
To get physically reasonable interpretation,
another distribution function as for example the $Q$--function
should be used.
The purpose of this Rapid  Communication is to
evaluate the phase information
included in the phase--sensitive data  using
 maximum likelihood estimation.  The phase
distribution  then yields the ultimate  resolution
 corresponding to the Fisher information.
As an explicit example, the quadrature measurement is interpreted
as  quantum phase measurement.
The  proposed method deals with the observed data
(\ref{data}) in the optimum way.

   Let us  formulate the problem for an arbitrary
multiple  measurement of  discrete
phase--sensitive observable \cite{H95}.
The case of quantum observable  with
continuous spectrum  will be obtained   by a straightforward
limiting procedure.
Assume the quantum measurement  of  quantum variable
$\hat Y$ yielding  discrete spectrum $ |y_k \rangle $ enumerated
for brevity by a (multi)index  $k.$
 The purpose of  phase detection  is to determine the
non--random c--number
displacement parameter $\theta $ in the given interval
 entering the  phase  displacement
transformation \cite{Hel74} of quantum state as
$
|\psi(\theta) \rangle  = e^{-i\theta \hat N} |\psi\rangle, $
$ \hat N $ being a Hermitian  operator.
The variable  $\theta$ represents  the {\em  true value}  of the phase
shift.
The  estimation on the interval $\theta \in [0,2 \pi)$  will be
considered   for concreteness.
The probability   of finding
the complex amplitude $y_k$  by performing the measurement
in transformed  quantum state $|\psi(\theta) \rangle $ is  given  by
quantum mechanics as
$$
p_k(\theta)  = |\langle \psi |e^{i\theta \hat N} | y_k \rangle|^{2}.
$$
 Knowing all these probabilities  in dependence
   on the induced phase shift, an unknown phase shift
may be inferred on the basis of  multiple
  output data   $y_1, y_2,\dots,  y_n.$
Following the approach used in the Ref. \cite{H95}, the
conditional  phase distribution of inferring  phase shift $\phi$
when $\theta$ is  true,  is given by
 the normalized likelihood
function  \cite{B92} as
\begin{equation}
P (\phi|\theta) =\frac{1} {C_n(\theta)}
 \left\{ \prod_k  [p_{k}(\phi)]^{p_k(\theta)}\right\}^n.
\label{red}
\end{equation}
 The normalization is
 $
C_n(\theta)=
\int_{0}^{2 \pi} d\phi \left\{\prod_k [ p_{k}(\phi)] ^{p_k(\theta)}
\right\}^n$ and index $k$ exhausts all the possible
 values appearing with nonzero probability.
 The number of samples  $n$ is assumed  to  be
sufficiently large in order to get statistically significant
sampling.
The likelihood function may be expressed using the
relative entropy
\begin{equation}
S(\phi|\theta) = - \sum_{k} p_k(\theta)\ln p_k(\phi)
\label{mut}
\end{equation}
as
\begin{equation}
P (\phi|\theta) \propto e^{- n S(\phi|\theta)}.
\label{Pe}
\end{equation}
The case of phase sensitive  observables  with
continuous spectrum $y$  may be easily incorporated in this step
defining the relative entropy as
\begin{equation}
S(\phi|\theta) = - \int dy\; p(y,\theta)\; \ln p(y,\phi) .
\label{cont}
\end{equation}
The preferred phase shift is given by
the true value $\theta$, since the relative entropy has minimum at
  $ S(\phi = \theta|\theta) $ due to the  Gibbs inequality
\cite{CD94}
\begin{equation}
 S(\phi|\theta) \ge S(\phi = \theta|\theta).
\end{equation}
The estimation is biased, but may be sometimes
  well  approximated  by  the Gaussian distribution
 with the variance predicted by the Fisher
information.   Using the Taylor decomposition of
$\ln p_k(\phi) $  at the point $\phi = \theta$  the
relative entropy     (\ref{mut}) reads
\begin{equation}
S(\phi|\theta) \approx - \sum_k \{ p_k(\theta)\ln p_k(\theta) -
\frac{1}{2} \frac{[p_k'(\theta)]^2}{p_k(\theta)} [\phi - \theta]^2 +
...\}.
\end{equation}
The prime denotes the  derivative
$p'_{k}(\theta) = {d p_k(\phi)} / {d \phi}|_{\phi = \theta}. $
The first term represents the Shannon entropy
$$ S(\theta) = - \sum_{k} p_k(\theta)\ln p_k(\theta), $$
whereas the second one is  the Fisher information
$$
 I(\theta) = \sum_k \frac {[p_k'(\theta)]^2}{p_k(\theta)}.
$$
The variance of  phase distribution in this approximation is
simply
$$  \Delta \phi  = \frac{1}{\sqrt{n I}} .$$
Provided that the   Gaussian  approximation  cannot be used,
the phase resolution
may be always  evaluated  using dispersion
$$
D(\theta) =\sqrt{ 1 - |\langle e^{i \phi} \rangle |^{2}}.
$$
Here the averaging over the phase is performed in the  specified
phase interval. Dispersion depends on the true value
of phase shift, in general. For sharp measurements it
corresponds to the  ordinary  notion of
 variance $  D \approx \Delta \phi $ restricted to the finite
interval \cite{Hel74,Rao73}.

 As an explicit example assume now the quantum measurement
 of phase--sensitive quadrature component (\ref{quadr}) performed for
concreteness   in  the  coherent
 state  with the complex   amplitude
$\alpha = |\alpha| e^{i\varphi}; $
$ |\psi \rangle  = \hat  D(\alpha) |0\rangle.$
 Here the  displacement   operator is
$ \hat D(\alpha) =  \exp(\alpha \hat
a^{\dagger} - \alpha^*\hat a). $
The phase shift of single--mode field is generated by the
photon--number operator   $  \hat N  = \hat a^{\dagger} \hat a. $
The probability of
finding the value $x$ of rotated quadrature operator
(\ref{quadr}) may be specified for the given signal
state as
\begin{equation}
p(x,\theta')  = \frac{1}{\sqrt{\pi}} \exp\{-
[x -\sqrt{2} |\alpha| \cos\theta']^2\},
\label{pex}
\end{equation}
where $\theta' = \theta-\varphi$  is the {\em phase difference }
between local oscillator and signal fields.
The quantum estimation problem is the following:
The  distribution  (\ref{pex}) is explicitly known  as a
function of  quadrature phase  difference $\theta'$
and quadrature component
$x.$  These dependencies are always experimentally measurable and
for example  may be  scanned in advance.
 The particular choice of Gaussian distribution
 represents an easy  example consistent
with the assumptions of Refs. \cite{VS91} and \cite{BSR93}.
  Using  this knowledge,  an {\em a priori} unknown  fixed phase
difference  should be inferred  as accurate as possible on the basis
of
 limited number of measured  data  $x_1, x_2, \dots, x_n.$
The straightforward application of the theory yields the
relative  entropy as
\begin{eqnarray}
\nonumber
S(\phi|\theta') = \frac{1}{2} \ln \pi  + \int^{\infty}_{- \infty}  dx \;
p(x,\theta')
 [ x -\sqrt{2} |\alpha|\cos\phi]^2
\\
\label{entropy}
\\
 \nonumber  =  \frac{1}{2} \ln \pi  +\frac{1}{2} + 2 |\alpha|^2
[\cos\phi   - \cos  \theta' ]^2,
\end{eqnarray}
where $\phi$ is estimated (inferred) phase difference.
The phase distribution inferred after n trials then  reads
\begin{equation}
P(\phi|\theta')  =  C(\alpha, \theta')
\exp \{- 2 n |\alpha|^2  [\cos\phi
  - \cos \theta' ]^2\},
\label{vysl}
\end{equation}
where the normalization is
$$
 C(\alpha, \theta')  = \{ \int_0^{2\pi} d\phi
 \exp \{- 2 n |\alpha|^2  [\cos\phi  - \cos \theta' ]^2 \} \}^{-1}.
$$
Significantly,  the resulting  phase  distribution is not
shift invariant  depending
only on the difference $\phi - \theta'.$
In general, the  phase estimation is biased.
It may be interpreted using
the standard definition of the normal distribution on the circle
(von Mises) \cite{Rao73}
\begin{equation}
f(x) = \frac{1}{2\pi I_0(\kappa)} \exp[\kappa \cos(x-\beta)],
\end{equation}
$x, \beta \in [0,2\pi), $ which is
centered at $\beta$ and  characterized by the dispersion
$D^2 = 1 - [I_1(\kappa)/I_0(\kappa) ]^2; $
$I_0(\kappa), I_1(\kappa)$ being the modified Bessel functions.
The phase distribution (\ref{vysl}) may be written as
 the normalized product of two von Mises distributions
\begin{equation}
P(\phi|\theta') \propto f_1(\phi) f_2(2\phi)
\end{equation}
centered at $\beta_1 = 0, \beta_2 = -\pi.  $
The parameters are
$\kappa_1 = 4 n |\alpha|^2  \cos\theta', \;\;\;
 \kappa_2 =  n |\alpha|^2  .$

Let us  detail  the phase  information  included in such a phase
measurement.
The phase distribution (\ref{vysl})  exhibits the mirror
symmetry
since
$ P(\phi|\theta') = P(2\pi - \phi|\theta').$
 Hence the phase
measurement  yields the one--peak  distribution  on the interval
$[0,2 \pi)$ only if $\theta' = 0$ or $\theta' = \pi.$
These two possibilities   are of course distinguishable  by the sign
of the measured quadrature components $x_i,$ as  the
probability distribution (\ref{pex})  indicates.
 Unfortunately, the phase measurement near the
points $\theta' = 0$ or $\theta' = \pi$ yields rather bad
resolution, as will be seen in the following.
In all the other cases   of phase differences $\theta'$, the
detected  phase sensitive data corresponding to the statistics
(\ref{pex}) do not distinguish between the values $\theta'$ and
$2\pi -\theta'$ and therefore neither the  inferred phase
distribution (\ref{vysl}) does.
 This ambiguity
of phase measurement may be avoided
 estimating  the phase
difference  on the half--width interval $[0, \pi)$  only.
Normalization should be changed  to the half value in comparison
to the multiplicative factor in  distribution  (\ref{vysl}).
The inferred
 phase distribution in dependence on the true phase shift $\theta$
is plotted in the  Fig. \ref{fig1}  for the input coherent field
with the real amplitude $(\varphi = 0).$
The estimated phase shift is always  localized around the true
value, but in general the phase estimation is biased.
The accuracy may be easily assessed for
an appropriately squeezed input  state characterized by
the wave function analogous to
(\ref{pex})
\begin{equation}
p_{sq}(x,\theta')  = \frac{e^r}{\sqrt{\pi}} \exp\{-  e^{2r}
[x -\sqrt{2} |\alpha| \cos\theta']^2\},
\label{pexsq}
\end{equation}
$r$ being the squeezed parameter. The value $r = 0$ corresponds to
the coherent state.

The phase  information is   the sharpest near the point
$ \theta'  =  \pi/2. $
The resolution  may be evaluated as $$ \Delta \phi|_{\theta' = \pi/2}
\propto \frac{e^{-r}}{\sqrt{ n} |\alpha|  }. $$
Assuming further the optimum partition of squeezed state   energy
$ N = |\alpha|^2 + \sinh^2 r $ as
$ e^r/2 \approx |\alpha| \approx
\sqrt{N/2},$
the optimum resolution corresponds to the well known
 ultimate limit of squeezed state  interferometry
${1}/{\sqrt{ n  N^2} }.  $
No optimization is necessary in the case of coherent state
 yielding  the limit of
coherent state  interferometry
${1}/{\sqrt{ nN }}.  $
This statistical analysis well corresponds to the semiclassical
(linear) approximation, when the phase resolution is predicted by
the  intrinsic fluctuations of the signal
$$  |\frac{d \langle X(\theta') \rangle }{d \theta'}| \; \Delta
\theta'
= \Delta  X(\theta').$$
Using the
distribution (\ref{pexsq}) we  conclude
$
\langle X(\theta') \rangle  = \sqrt{2} |\alpha| \cos\theta',
\;\;
\Delta X(\theta') = e^{-r}/\sqrt{2}  $  and  therefore
$$\Delta \theta \propto \frac{e^{-r}}{|\alpha| |\sin\theta'|}.$$
Semiclassical treatment  represents
 good  estimation in the regime of the best
resolution, nevertheless it  failures at the points close to
 $\theta' = 0.$ Moreover, here  also
the Fisher information tends to zero, since the  quadratic term
in the relative entropy  (\ref{entropy}) disappears.
The necessary assumptions concerning the existence of the
 Fisher information are not
fulfilled   and, for example, the Cram\'{e}r--Rao bound is  not
valid \cite{B}.
Nevertheless, the analysis of the phase distribution (\ref{vysl})
shows that the phase resolution reads
$$
\Delta \phi|_{\theta' = 0} \propto \sqrt{ \Delta \phi|_{\theta' = \pi/2
}},$$
 yielding considerably worse phase resolution at this point.

The block diagram of the phase detection based on the maximum
likelihood estimation is sketched in the Fig. \ref{fig2}.
Assuming the
phases of the local oscillator and input coherent states as
 $\theta$ and  $\varphi,$
the balanced homodyne detection  measures the statistics of the
quadrature operator  (\ref{quadr}) at the point $\theta +\pi/2.$
The phase difference may be estimated
using the likelihood function of measured phase sensitive data
yielding the conditional phase distribution
\begin{equation}
P_{homod}(\phi|\theta') \propto \exp[ - 2 n |\alpha|^2( \sin \phi -
\sin\theta')^2].
\label{vhom}
\end{equation}
  The predicted  phase  resolution  is   as in the Fig.
\ref{fig1}, but
shifted by the value $\pi/2$ in both the true and inferred
phases. The best resolution is  then achieved if $\theta - \varphi
 = 0. $  For  coherent input field the   phase distribution  at
this point  reads
$$
P(\phi|\theta'= 0) \propto \exp( - 2 n |\alpha|^2  \sin^2 \phi).
$$
Assuming further the total energy needed for such a realization
of multiple measurement as $n|\alpha|^2, $ the phase distribution
may be compared with the  proposal of Vogel and Schleich \cite{VS91}.
The relations  (\ref{vo}) and  (\ref{pex}) at
the phase $\theta +\pi/2$  yield   in this particular case
the phase distribution predicted by the maximum likelihood   estimation.
 Nevertheless, it need not be  necessarily so
in the general case of detection of an arbitrary phase shift,
using an arbitrary input state or using an arbitrary
 phase sensitive variable. Estimation theory
therefore naturally extends the ``phase without phase concept".

We demonstrated that any phase--sensitive measurement may serve
for  statistical  prediction  of phase shift. The content of phase
information may be evaluated using the relative entropy of the
phase in dependence  on the observable probabilities  only.
The resolution predicted by the Fisher information is achieved
if it exists. The proposed method
based on the maximum likelihood  estimation uses the information
accumulated in the process of multiple measurement  in the optimum way.
This treatment better suits to the experimental  conditions than
sophisticated phase concepts. Particularly, the phase
distribution obtained in the process of measurement is rather associated
with nonlinear functionals (likelihood  functional, relative
entropy) than  with the linear ones such as
   the  distribution functions  on the phase space  are
(Wigner function, Q--function).
Moreover, since the phase distribution in realistic experiments
is biased  and  phase shift dependent, the   detailed statistical
analysis free of any a priori restricting assumptions
 is always necessary.

This contribution was partially supported by the internal grant
of Palacky University.

\begin{figure}
\caption {Phase distribution as function of inferred phase shift
$\phi$ in dependence
on the true phase shift $\theta$   for
coherent input with total energy  $ n |\alpha|^2  = 100.$ }
\label{fig1}
\end{figure}

\begin{figure}
\caption {
Scheme of homodyne detection used for phase difference
measurement.}
\label{fig2}
\end{figure}

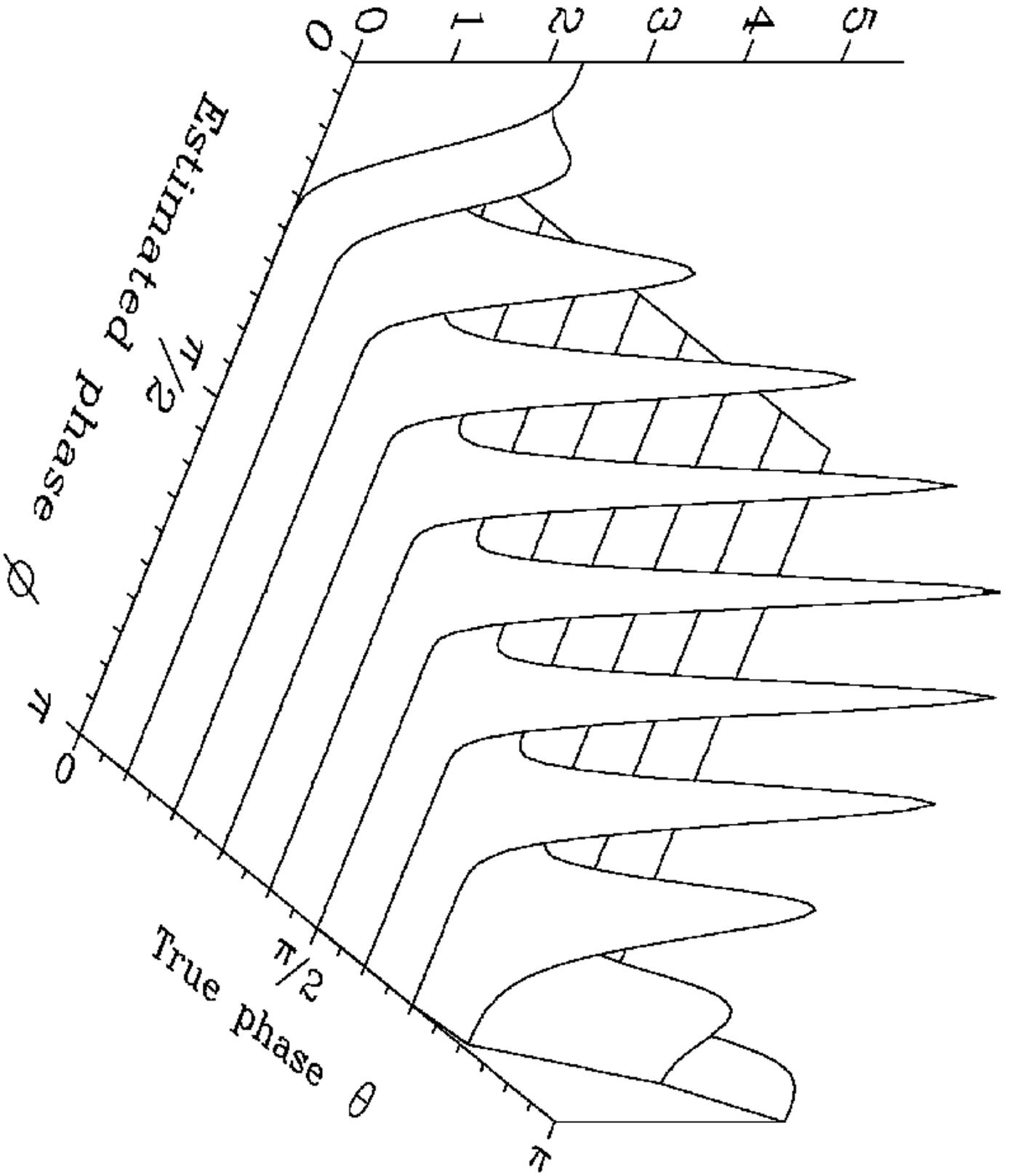
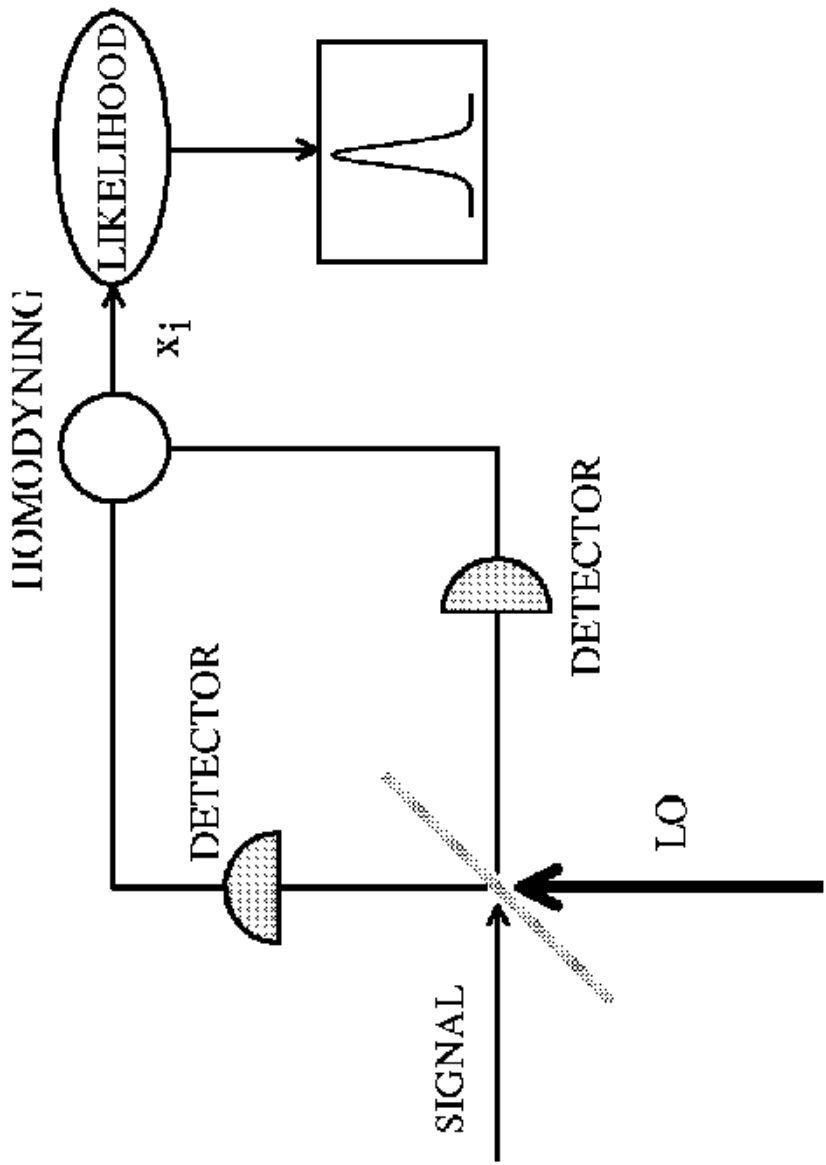
\end{document}